\documentclass[pra,twocolumn,longbibliography]{revtex4}

\usepackage[utf8]{inputenc}
\usepackage{bm}
\usepackage{mathtools}
\usepackage{bbold}
\usepackage{amsfonts, amsmath, amsthm, amssymb}
\usepackage{hyperref}
\usepackage[normalem]{ulem}
\usepackage{color}
\usepackage{braket}
\usepackage{float}
\usepackage[dvipsnames]{xcolor}
\usepackage{graphicx,subfigure}

\usepackage{subfiles}

\begin{document}

\title{
Non-local quench spectroscopy of fermionic excitations in 1D quantum spin chains
}
\author{Saverio Bocini}
 \email{saverio.bocini@ens-lyon.fr}
\affiliation{%
Univ Lyon, Ens de Lyon, CNRS, Laboratoire de Physique, F-69342 Lyon, France
}%
\author{Filippo Caleca}
\affiliation{%
Univ Lyon, Ens de Lyon, CNRS, Laboratoire de Physique, F-69342 Lyon, France
}%
\author{Fabio Mezzacapo}
\affiliation{%
Univ Lyon, Ens de Lyon, CNRS, Laboratoire de Physique, F-69342 Lyon, France
}%
\author{Tommaso Roscilde}
\affiliation{%
Univ Lyon, Ens de Lyon, CNRS, Laboratoire de Physique, F-69342 Lyon, France
}%

\begin{abstract}
The elementary excitations of quantum spin systems have generally the nature of weakly interacting bosonic quasi-particles,  generated by local operators acting on the ground state. Nonetheless in one spatial dimension the nature of the quasiparticles can change radically, since many relevant one-dimensional $S=1/2$ Hamiltonians can be exactly mapped  onto models of spinless fermions with local hopping and interactions. Due to the non-local nature of the spin-to-fermion mapping, observing directly the fermionic quasiparticle excitations is impossible using local probes, which are at the basis of all the forms of spectroscopy (such as neutron scattering) traditionally available in condensed matter physics. Here we show theoretically that \emph{quench spectroscopy} for synthetic quantum matter -- which probes the excitation spectrum of a system by monitoring the nonequilibrium dynamics of its correlation functions -- can reconstruct accurately the dispersion relation of fermionic quasiparticles in spin chains. This possibility relies on the ability of quantum simulation experiments to measure non-local spin-spin correlation functions, corresponding to elementary fermionic correlation functions.  Our analysis is based on new exact results for the quench dynamics of quantum spin chains; and it opens the path to probe arbitrary quasiparticle excitations in synthetic quantum matter.   
\end{abstract}

\maketitle

\paragraph*{Introduction.} A distinctive feature of the various phases of quantum matter is their excitation spectrum at low energy, which is typically a direct reflection of the nature of the ground state \cite{Sachdev-book}. In systems displaying long-range order in the ground state, the elementary excitations can be thought of as harmonic fluctuations around a classically ordered ground-state configuration, akin to those of a crystalline solid -- and as such they have the nature of bosonic quasi-particles, created by local operators. Such is the case of ordered quantum magnets, whose elementary excitations are harmonic spin waves, which can be generated by acting with a spin-flip operator (or, more precisely, with its Fourier transform) on the ground state. As a consequence, the dispersion relation of these elementary excitations is immediately accessible to probes that couple \emph{locally} with the system of interest -- such as \emph{e.g.} neutron scattering \cite{Boothroyd-book,Blundell2001}, whose cross section is related to the Fourier transform of a two-time/two-site spin-spin correlation function $\langle \sigma_{\bm q}^\mu(t) \sigma_{-\bm q}^\mu(0)\rangle$, where $\sigma_{\bm q}^\mu = \frac{1}{\sqrt{V}} \sum_i e^{i\bm q \cdot \bm r_i} \sigma_i^\mu$ ($\mu= x,y,z$) is the Fourier transform of Pauli matrices (for $S=1/2$ systems) at the sites $i$ of a lattice of volume $V$, and $\langle ... \rangle$ is the expectation value on the equilibrium state of the system.  Other forms of spectroscopy in condensed matter (such as nuclear magnetic resonance \cite{Slichter-book}, muon-rotation spectroscopy \cite{Muon-book}, etc. ) are also based on local probes.  On the other hand, the phases of matter which depart most radically from classical order at low energies can host excitations differing fundamentally from the above picture of wavelike deviations from the ground state. This is the case \emph{e.g.} of fractionalized excitations in spin liquids \cite{Sachdev-book, Savary2017} related to those (photons, magnetic/electric charges) of lattice-gauge theories; or of the fermionic excitations of quantum spin chains, which are the focus of this work. These excitations might have the nature of well-defined quasi-particles, but they manifest themselves in a weak form (namely as continua of excitation frequencies) \cite{Lake2005,Coldea2011,Han2012,Laurell2021} to local spectroscopies as those cited above, since they are unable to excite and/or probe them in a selective manner. 

In this context experiments on quantum many-body devices -- based \emph{e.g.} on ultracold neutral atoms \cite{Browaeys2020}, trapped ions \cite{Monroe_2021} or superconducting circuits \cite{Garcia-Ripoll-book} -- not only realize a form of ``synthetic" quantum matter, but they open new possibilities for the spectroscopy of elementary excitations, since they can give microscopic access to the individual degrees of freedom. This allows in principle for the measurement of arbitrary observables, accompanied by the ability of driving the system arbitrarily far from equilibrium. 
Based on this unprecedented capabilities, the concept of \emph{quench spectroscopy} has been recently put forward theoretically \cite{MenuR2018, Villa2019, Villa2020, Menu2023} and demonstrated experimentally in Rydberg-atom arrays \cite{chen2023spectroscopy}. Quench spectroscopy is based on monitoring the time evolution of (equal-time) correlation functions, such as $\langle \psi(0) | \sigma_i^\mu(t) \sigma_j^\mu(t) |\psi(0)\rangle$ -- where $|\psi(0)\rangle$ is the initial state of a non-equilibrium unitary evolution. A time-space Fourier analysis of the above correlation function gives insight into the dispersion relation of the elementary excitations. In this work we extend the above concept to that of \emph{non-local} quench spectroscopy (NLQS), based on the measurement of the time dependence of non-local correlation functions, i.e. correlations between strings of operators. Making use of new exact results on the non-equilibrium dynamics of quantum spin chains, we show that NLQS can directly probe fermionic quasiparticles which represent their truly elementary excitations.

\paragraph*{Quantum spin chains and mapping to free fermions.} 

We focus our attention on one-dimensional $S=1/2$ spin models defined on a lattice of $L$ sites with label $i=1,..,L$. Spin-1/2 operators associated with Pauli matrices $\sigma_i^\mu$ can be generally mapped onto fermions via the Jordan-Wigner (JW) transformation
\cite{Jordan:1928wi,LSM1961,Franchini_2017}
$\sigma_i^+ = \mathcal{O}_{1,i-1} \ c_i^\dag$,
$\sigma_i^z = 2c_i^\dag c_i - 1$, 
where $c_i^\dag$, $c_i$ are fermionic creation and annihilation
operators and 
$\mathcal{O}_{i,j}=\prod_{l=i}^{j}\sigma_l^z$ is the so-called string operator.  In this work we specialize to the class of systems described by the XY model in a transverse field
\begin{equation}
    H = J\sum_{j=1}^{L} 
    \left(
    \frac{1+\gamma}{2}\sigma^x_{j}\sigma^x_{j+1}+
    \frac{1-\gamma}{2}\sigma^y_{j}\sigma^y_{j+1}+
    h \sigma^z_j
    \right)~.
    \label{eq:H_spinXY}
\end{equation}
In the rest of this work we consider chains with an even length $L$ and with periodic boundary conditions ($\sigma^\alpha_{L+1}\equiv \sigma^\alpha_{1}$) are assumed; $h$ is the transverse magnetic field, and the parameter $\gamma$ governs the anisotropy of interactions in the $xy$ plane.
 
 The JW transformation maps Eq.~\eqref{eq:H_spinXY}  onto a quadratic fermionic Hamiltonian, up to a constant  \cite{LSM1961,Franchini_2017}:
\begin{align}
    H = \ & J \sum_{j=1}^{L-1} \bigg(
    c_j c^\dag_{j+1}+ \gamma c_j c_{j+1} \ 
    + \text{h.c} \bigg) + 2Jh\sum_{j=1}^{L} c^\dag_jc_j  \nonumber \\
    & -J \mathcal{O}_{1,L} \
    \bigg( c_L c^\dag_1 + \gamma 
    c_L  c_1 + \text{h.c.}\bigg)~.
    \end{align}
The boundary terms on the second line 
are negligible in the thermodynamic limit, but must be taken
into account for finite-size systems. 
As the Hamiltonian only creates or annihilates pairs of fermions,
fermion-number parity is conserved, namely
$[\mathcal{O}_{1,L},H]=0$.
This implies that the Hamiltonian eigenspectrum can be separated into
two different parity sectors with $\mathcal{O}_{1,L}=\pm1$,
which is equivalent to considering periodic (for ${O}_{L}=+1$) or antiperiodic (for ${O}_{L}=-1$) 
boundary conditions for the fermions \cite{Franchini_2017}.
The Hamiltonian can then be written as  $\sum_{\Gamma=\pm} P^\Gamma H^\Gamma$,
where $P^\pm$ is the projector onto the 
parity sector $\mathcal{O}_{1,L}=\pm1$ and $H^\pm$ the corresponding Hamiltonian. 
Within each parity sector, $H^\pm$ can be diagonalized by moving to fermionic operators in momentum space 
$c_{k_{\pm}}=\sum_l e^{-ik_{\pm }l} c_l / \sqrt{L}$ where $k_+ = k$, $k_-  = k + \pi/L$, and $k = 2\pi n/L$ and $n = 0, ... ,L-1$.
As detailed in the Supplemental Material (SM) \cite{SM}, the Bogoliubov-rotated operators 
 $\eta_k = \cos(\theta_{k_\pm}/2) \ c_{k_\pm} - \sin(\theta_{k_\pm}/2) \ c^\dag_{-k_\pm}$ 
with $\epsilon_k \sin(\theta_k)=-2J\gamma\sin(k)$ and $
 \epsilon_k \cos(\theta_k)=2J(h-\cos(k))$,
lead to the diagonal form for the Hamiltonian
\begin{equation}
    H^{\pm} =  \sum_{k} \epsilon_{k_{\pm}}
    \bigg( \eta^\dag_k \eta_k - \frac{1}{2} \bigg) \label{e.Hpm}
\end{equation}
with $\epsilon_{k}=2|J|\sqrt{(h-\cos k)^2+\gamma^2\sin^2k}$.

\emph{Quench spectroscopy of elementary excitations.} The XY chain in a transverse field has been realized in several experiments, going from condensed matter \cite{Coldea2011, Laurell2021} to synthetic quantum matter based on neutral atoms \cite{Simon2011,Labuhn2016, Bernien2017}, trapped ions \cite{Monroe_2021} and superconducting circuits \cite{King2022}, to cite a few examples. Nonetheless the sharp spectrum of its fermionic excitations has so far eluded a direct experimental observation. 
In this context, quench spectroscopy  (QS) \cite{MenuR2018, Villa2019, Villa2020, Menu2023} offers a very promising alternative, but so far it  has been investigated theoretically \cite{MenuR2018, Villa2019, Villa2020, Menu2023} as well as experimentally \cite{chen2023spectroscopy} only in relationship with the evolution of  correlation functions of local operators, more precisely of their Fourier transform (FT), starting from a given initial state 
$|\psi(0)\rangle$. Defining local observables $\{A_j\}$ (such as local spin operators $A_j = \sigma_j^\mu$),  and their Fourier transform $A_k = \sum_j e^{-ikj} A_j /\sqrt{L}$, QS consists in probing the quench spectral function (QSF) \cite{MenuR2018, Villa2019} $Q_A(k,\omega) = \sum_{nm} \langle \psi(0) | n \rangle \langle m |\psi(0)\rangle \langle n | A_k A_{-k} | m \rangle \delta (\omega - \omega_{nm})$,
 which is the temporal FT of the time-dependent structure factor $S_A(k,t) = \langle A_k A_{-k} \rangle (t)$ -- the latter being the spatial FT of the correlation function $C_A(l,t) = \langle A_i A_{i+l} \rangle(t)$. Here $|n\rangle, |m\rangle$ are Hamiltonian eigenvectors with energies $E_n, E_m$, and $\omega_{nm} = E_n - E_m$ (assuming $\hbar = 1$). Hence the QSF probes transitions between Hamiltonian eigenstates that have significant overlap with the initial state; and which are connected by the creation of two excitations of momenta $k$ and $-k$ generated by the $A_{\pm k}$ operators. 

 Considering \emph{e.g.} the XY chain of Eq.~\eqref{eq:H_spinXY} in zero field ($h = 0$), the operator $A_k = \sigma_k^z$ generates wavelike spin flips along the axis (or axes) of strongest spin correlations, lying in the $xy$ plane. It is then natural to monitor the evolution of the correlation function 
$C_{\sigma^z}(l,t)=\langle \sigma_i^z \sigma_{i+l}^z \rangle(t)$ and of the related time-dependent structure factor $S_{\sigma^z}(k,t) = \langle \sigma_k^z \sigma_{-k}^z\rangle(t)$, as done in recent experiments on two-dimensional XY models \cite{chen2023spectroscopy}. Yet, in the fermionic language, which is most appropriate for spin chains, the $\sigma_k^z$ operator generates particle-hole \emph{pairs} (for $\eta$ fermions) with a momentum transfer of $-k$; or pairs of $\eta$ particles/holes with momenta summing up to $-k$  \cite{SM}. Hence the QSF $Q_{\sigma^z}$ reveals \emph{e.g.} all possible double particle-hole excitations with opposite momentum transfers $\pm k$, or double pairs of particles/holes, which connect Hamiltonian eigenstates overlapping with the initial state. It is clear that many frequencies $\omega_{nm}$ correspond to the same wavevector $k$ in this case -- \emph{e.g.} $\omega_{nm} = \epsilon_{-k+q_1} + \epsilon_{k+q_2} - \epsilon_{q_1} -\epsilon_{q_2}$ for double particle-hole pairs (with $q_1$ and $q_2$ arbitrary wavevectors). These frequencies are all different  given that the dispersion relation $\epsilon_k$ is not linear in $k$. Hence the QSF $Q_{\sigma^z}$ will generically reveal a \emph{continuum} of excitation frequencies, similar to the continua revealed \emph{e.g.} by neutron scattering on spin chains \cite{Lake2005,Coldea2011,Han2012,Laurell2021}; and not the sharply defined frequencies of the fermionic quasiparticles. 

 \emph{Non-local quench spectroscopy.} Unlike condensed-matter spectroscopies, QS can be generalized to arbitrary correlation functions, i.e. to arbitrary operators $A_k$, by taking advantage of the individual addressing of the microscopic degrees of freedom offered by many platforms of synthetic quantum matter \cite{Browaeys2020,Gross2021,Monroe_2021}. When willing to probe fermionic excitations directly, the most natural choice is to consider \emph{fermionic} $A_k$ operators (instead of spin ones), namely to consider $A_k = c_k, c_k^\dagger$, so that the transitions probed by the associated QSF are actually associated with the creation/destruction of pairs of particles (or holes) with sharply defined momentum $k$. This amounts to monitoring the fermionic correlation function 
 \begin{align}
    C_F(\ell,t) & = \braket{c_1c_{\ell+1}+c^\dag_{\ell+1} c^\dag_1} \nonumber
    \\ & = \frac{1}{2}
    \braket{
    \big(\sigma^y_1 \sigma^y_{\ell+1} - \sigma^x_1 \sigma^x_{\ell+1} \big)
    \mathcal{O}_{2,\ell}}, \label{eq:C_F}
\end{align}
defined for $\ell\in\{0,...,L-1\}$,
which is non-local in the spins, since it looks at two sites at distance $\ell$ \emph{and} at all the sites in between via the string operator $\mathcal{O}_{1,\ell}$. The corresponding time-dependent structure factor is $S_F(k,t) = \langle c_k c_{-k} + c_{-k}^\dagger c^\dagger_k \rangle(t)$, whose temporal FT gives the fermionic QSF $Q_F(k,\omega)$~\cite{SM}. At finite frequency for generic Bogolyubov angles $\theta_k$, and at any frequency when $\theta_k\in\{0,\pi\}$, the fermionic QSF probes the transitions induced by the creation of two $\eta$ particles (or two $\eta$ holes) with opposite momenta $k$ and $-k$ \cite{SM}.  In the thermodynamic limit, in which the two fermionic parity sectors coincide, this gives
\begin{equation}
\begin{split}
    Q_F(k,\omega) = 2\pi F(k)
    \bigg(
    \delta(\omega-2\epsilon_k) +
    \delta(\omega+2\epsilon_k)
    \bigg),
\end{split}
\label{eq:SF_TDlim}
\end{equation}
where $F(k)=-\cos(\frac{\theta_k-\theta_{-k}}{2})\lim_{L\rightarrow\infty}\text{Im}\braket{ \eta_k \eta_{-k} }_0$, the expectation value is computed in the initial state, and we have used the fact that $\epsilon_k = \epsilon_{-k}$.
This result shows that NLQS based on monitoring the fermionic correlations in Eq.~\eqref{eq:C_F} gives direct access to the dispersion relation of the fermionic quasiparticles. This ability rests fundamentally upon preparing initial states $|\psi(0)\rangle$ that possess two-particle/two-hole coherences $F(k)$: while this is clearly impossible in systems of  real fermionic particles, it is instead rather natural for fermionic quasiparticles onto which quantum spins are mapped. Finite size brings corrections to Eq.~\eqref{eq:SF_TDlim}, due to the coexistence of two sectors in the fermionic theory of Eq.~\eqref{e.Hpm}, as we will show in details in the examples below. 

\begin{figure}
    \centering
    \includegraphics[width=.95\columnwidth]{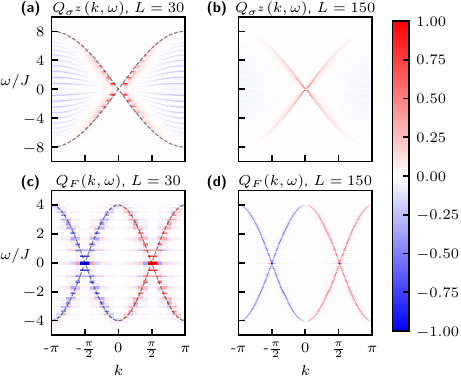}
    \caption{\emph{Quench spectral functions for the XX chain.} (a-b) Spin QSF $Q_{\sigma^z}(k,\omega)$ for to system sizes, $L=30, \ 150$;
    (c-d) Fermionic QSF $Q_F(k,\omega)$  for the same sizes. The dashes gray lines in (a) and (c) correspond to $\pm 8J\sin(k/2)$ and $\pm 4J \cos(k)$ respectively. In all panels the time Fourier transform used the quench evolution over the interval  $T|J|=L/3$; and spectral weights have been renormalized to their maximum absolute values.}
    \label{fig:Stilde_XXchain}
\end{figure}

\paragraph*{XX chain.}
The XX chain is defined as the Hamiltonian of Eq.~\eqref{eq:H_spinXY} for $\gamma, h=0$. 
In this case the Bogoliubov angle is trivial and Jordan-Wigner transformation plus FT already bring the spin Hamiltonian into the diagonal form Eq.~\eqref{e.Hpm} with 
$\epsilon_k = |2J\cos(k)|$ \footnote{A particle hole transformation on the inner half of the Brillouin zone reconstructs then the usual free-particle dispersion relation $-2J \cos(k)$.}. To reconstruct this dispersion relation via NLQS, one can take as initial state 
the coherent spin state (CSS) along $x$, i.e. $\ket{\psi(0)}=\ket{\text{CSS}_x}=
\otimes_i \ket{\rightarrow_x}_i$, which, in the fermionic language, corresponds to a linear combination of two Gaussian states with equal weights \cite{SM}. The fermionic time-dependent structure factor can be exactly 
computed with free fermions techniques \cite{SM} to give the explicit form
\begin{align}
     &S_F(k,t) = \frac{1}{2}\cos\big(4tJ\cos(k)\big)\sin(k) \ + \\ 
    &\frac{1}{2L}\sum_{k_{-}} (-1)^{k_-} \ \mathcal{F}(k-k_-) \sin\big(k_-+4tJ\cos (k_-)\big) \nonumber
    \label{eq:S_k_t_XX}
\end{align}
where $\mathcal{F}(x) = \cot(x)$ for $L/2$ odd and $\csc(x)$ for $L/2$ even; and $(-1)^{k_-} = (-1)^n$ with $k_- = (2n+1)\pi/L$.  This formula clearly delineates the contributions from the two fermionic sectors: the first line (resp.~second) is the contribution from the PBC (resp.~APBC) sector; as $L \to \infty$ the two contributions become equivalent.
    
\begin{figure*}
    \includegraphics[width=0.8\textwidth]{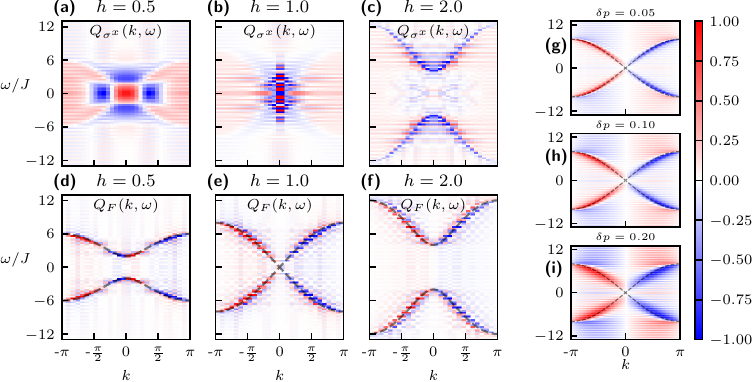}
    \caption{\emph{Quench spectral functions for the quantum Ising chain.} 
    (a-c) Spin QSF $Q_{\sigma^x}(k,\omega)$ for three values of the transverse field $h = 0.5, 1$ and 2; (d-f) Fermionic QSF $Q_F(k,\omega)$ for the same field values. Dashed lines indicate the dispersion relation of fermionic quasiparticles; (g-h) Fermionic QSF at the critical point $h=1$ including read-out errors with probability $\delta p$ (=0.05, 0.1 and 0.2). All conventions for the Fourier transform and the normalization of the spectral weights are as in Fig.~\ref{fig:Stilde_XXchain}.}
    \label{fig:Stilde_ising}
\end{figure*}


    We monitor the evolution of $S_F(k,t)$ for 
a time $T|J|=L/3$ in order to have a frequency resolution $2\pi/T$ scaling with system size.
A numerical FT in time reconstructs the fermionic QSF $Q_F(k,\omega)$. The latter is shown in  
Fig.~\ref{fig:Stilde_XXchain}, and compared with the QSF $Q_{\sigma^z}(k,\omega)$ from spin-spin correlations \cite{chen2023spectroscopy}.
We can clearly see that the spin QSF reveals a continuum of excitations, as anticipated previously. The continuum becomes very sharply peaked at the frequencies $\omega_k = \pm 8J\sin(k/2)$ when increasing the system size. These frequencies correspond to the edges of the particle-hole continuum, namely to $\pm 2 \max_q \Delta_{k,q}$ with $\Delta_{k,q} =|\epsilon_{k+q} - \epsilon_{k}|$, which give rise to a stronger signal among all particle-hole excitations at wavector $k$ because of the multiplicity of wavevectors $q$ which nearly maximize the function $\Delta_{k,q}$.  On the other hand the fermionic QSF shows clearly the dispersion relation of the fermions, more sharply so the bigger the system size. Indeed for a finite $L$ the coexistence of the two fermionic sectors leads to several frequencies associated to each wavevector \cite{SM}, with the dominant ones differing by terms of $O(1/L)$, such as $2|\epsilon_{k_+} - \epsilon_{k_-}| = |4J \pi \sin(k) |/L + O(L^{-2})$. Hence this implies an intrinsic spectral broadening of order $O(1/L)$, most prominent for $k = \pm \pi/2$, and vanishing when $k \to 0$ and $k \to \pm \pi$.


\paragraph*{Quantum Ising chain.} 
A second most relevant limit of the Hamiltonian of Eq.~\eqref{eq:H_spinXY} is given by the Ising chain in a transverse field, realized for $\gamma=1$ and a finite $h$. The ground state of this model exhibits a well-known quantum phase transition \cite{PFEUTY1970} for $h = 1$, separating two gapped phases -- a paramagnetic one for $h>1$, and a ferromagnetic one for $h<1$. In the paramagnetic phase the spins are strongly polarized by the field along the $z$ axis, so that in this case the $\sigma^{x(y)}_k$ operators are the ones generating wavelike spin flips with respect to a classical reference configuration. On the other hand, in the ferromagnetic phase long-range correlations develop along the $x$ axis, so that the  $\sigma^{z(y)}_k$ operators appear to be the most appropriate ones to generate wavelike spin flips around a classical approximation of the ground state.  
Here we consider a quench from the initial state polarized along the field axis $\ket{\text{CSS}_z} = \otimes_i | \uparrow_z \rangle_i$, which is a low-energy state for $h \gtrsim 1$ -- see the SM \cite{SM} for a study of the quench starting from $\ket{\text{CSS}_x}$. Fig.~\ref{fig:Stilde_ising}(a) shows that, at sufficiently high field $h$, the spin QSF $Q_{\sigma^x}$ reproduces the dispersion relation of the elementary excitations, in spite of their intrinsic fermionic nature. This is due to the fact that for large $h$ the excitations triggered by the quench form a dilute gas, so that their fermionic statistics is not relevant. Indeed the large-$h$ phase of the quantum Ising chain can be well described via spin-wave theory, which represents the excitations as bosonic \cite{Schneider2021,Menu2023}.  Nonetheless, upon lowering the field $h$ down to the quantum critical point $h=1$ the spin QSF acquires a broad structure, revealing again a continuum of frequencies for each wavevector (Fig.~\ref{fig:Stilde_ising}(b)); and this situation persists for even lower fields (Fig.~\ref{fig:Stilde_ising}(c)). The same applies when considering the other spin QSF, namely $Q_{\sigma^z}$, in the quench from the $\ket{\text{CSS}_x}$ state. 
On the other hand, the fermionic QSF allows one to reconstruct the quasi-particle dispersion relation for \emph{any} value of the field $h$, as shown in Fig.\ref{fig:Stilde_ising}(b). Most remarkably, one can perfectly well reconstruct the closing and reopening of the excitation gap in the spectrum upon lowering $h$ across the its critical value. This implies that the non-equilibrium dynamics of the quantum Ising chain can reveal perfectly its ground-state quantum phase transition.

\paragraph*{Experimental relevance and errors in parity measurements.}

From the experimental point of view, many recent experiments on synthetic quantum matter allow for the measurement of local spin components, with a different local measurement basis for each spin. This has been clearly demonstrated in experiments on trapped ions \cite{Brydges2019} and superconducting circuits \cite{Satzinger2021} in the context of randomized-measurement protocols \cite{Elben2023}; more recently the same ability has also been demonstrated in Rydberg-atom arrays \cite{Bornetetal2024}. These advances completely pave the way for the measurement of the fermionic correlation function in Eq.~\eqref{eq:C_F}. Similar string correlation functions have been measured in the past in quantum-gas microscopes \cite{Endresetal2011}. A legitimate question to ask concerns the robustness of the measurement of string correlation functions to the presence of detection errors. Indeed the operator $O_{\ell}$ contained in Eq.~\eqref{eq:C_F} is a parity, and parities are subject to significant errors if the readout fidelities of the individual qubits is not perfect. Yet, as we will see, the reconstruction of the quasi-particle dispersion relation via NLQS is significantly robust to detection errors. As commonly done in the literature \cite{chen2023spectroscopy,Bornet_2023,
PhysRevLett.121.123603}, we shall model the readout errors as independent qubit by qubit --  namely we consider that with a probability $\delta p$ the $\sigma_i^z$ operator at size $i$ is incorrectly measured, namely the recorded measurement is opposite to the state into which the qubit collapsed. Under this assumption, the probability of having $n$ incorrect parity measurements when dealing with a string of $l$ qubits is given by the binomial distribution
$
    p(n,l) = \binom{l}{n} \ \delta p^n \ (1-\delta p)^{l-n}.
$
Parities are affected only by an \emph{odd} number of readout errors; therefore the probability of an incorrect measurement is given by
$
    \mathcal{P}(l) = \sum_{n \ \text{odd}} p(n,l) = [1 - (1-2 \delta p)^l]/2~.
$
As string measurement outcomes are binary, in the presence of readout errors one should
correct the fermionic correlation function as $C^{(\delta p)}_F(l,t) = C_F(l,t)  (1-2\mathcal{P}(l-1)) = C_F(l,t) \exp(-l/\xi_{\delta p})$ where $\xi_{\delta p} = 1/|\log(1- 2\delta p)|$. Namely measurement errors simply introduce an additional exponential decay to the fermionic correlation function, with a characteristic decay length $\xi_{\delta p}$ diverging as $(2\delta p)^{-1}$ for $\delta p \to 0$. When taking the spatial FT, this length introduces a broadening $\delta k \approx 4\pi\delta p$  in $k$ space. Hence, if $\xi_{\delta p} \gtrsim L$ (in practice, if $\delta p \lesssim 1/(2L)$ for $L \gg 1$), then $\delta k \lesssim 2\pi/L$ (the natural finite-size resolution on momenta), and the readout errors affect minimally the experimental reconstruction of the fermionic QSF. 
Fig.~\ref{fig:Stilde_ising} (g)--(i) shows the fermionic QSF for the critical Ising chain for various values of $\delta p$, all exceeding $1/(2L)$; the gapless fermionic dispersion relation appears clearly in the QSF even for $\delta p$ as large as 20$\%$.

\paragraph*{Conclusions.}
In this work we have proposed non-local quench spectroscopy, which has the ability to reconstruct the dispersion relation of fermionic quasiparticles in quantum spin chains by monitoring the evolution of an arbitrarily non-local correlation function. Our proposal fully exploits the ability of recent experiments on synthetic quantum matter (neutral atoms, trapped ions, superconducting circuits, among others) to address individually their degrees of freedom, and thereby reveal the hidden nature of their elementary excitations; and it is robust to realistic read-out errors. This non-local form of spectroscopy would be completely unequaled in condensed matter, where all forms of spectroscopy are based on local probes. Moving beyond the case of free fermionic quasiparticles considered in this work, non-local quench spectroscopy can reveal the nature of elementary excitations when interactions among fermionic quasiparticles are present -- which is the case for generic quantum spin chains. And it can be generalized to any quantum system whose elementary excitations are defined in terms of any non-local operators. This is \emph{e.g.} the case of matter or gauge excitations in lattice gauge theories \cite{Kogut1979}, whose realization in synthetic quantum matter has been intensely discussed in the recent literature \cite{Aidelsburger2022,halimeh2023coldatomquantumsimulatorsgauge}. 

\begin{acknowledgments}
\emph{Acknowledgements.} This work was supported by the PEPR-Q project ``QubitAF''. Useful discussions with G. Bornet, A. Browaeys, C. Chen, P. Delplace, G. Emperauger, and T. Lahaye are gratefully acknowledged.
\end{acknowledgments}

\newpage
\null
\newpage

\section*{SUPPLEMENTAL MATERIAL}
\noindent{\bf \emph{Non-local quench spectroscopy of fermionic excitations in 1D quantum spin chains}
\section{From spins to fermions}}

We consider the same Hamiltonian of the main text and we transform it according to the Jordan-Wigner transformation
\begin{equation}
	\sigma^z_\ell=-i a_{2\ell-1} a_{2\ell}\,,
	\quad
	\sigma^x_\ell=\mathcal{O}_{1,\ell-1} a_{2\ell-1}\,,
\end{equation}
where the operators $a_j$ are Majorana fermions and satisfy the Majorana anti-commutation relations $a_i a_j+ a_j a_i=2\delta_{i,j}$, and $\mathcal{O}_{i,j}:=\prod_{l=i}^j\sigma^z_l$.
These Majorana fermions are related to the Jordan-Wigner fermions used in the main text by $c_l=\frac{a_{2l-1}-ia_{2l}}{2}$.
The Hamiltonian becomes
\begin{equation}
	\label{eq:generic_spin_ham_after_JW}
	H=\frac{1-\mathcal{O}_{1,L}}{2} H^+  
	+
	\frac{1+\mathcal{O}_{1,L}}{2} H^- \,,
\end{equation}
where $ H^+$ and $ H^-$ are two quadratic fermionic Hamiltonians
\begin{equation}
	H^{\pm}\equiv\frac{1}{4}\sum_{m,n=1}^{2L}  a_m \mathcal{H}^{\pm}_{m,n}  a_n \,,
\end{equation}
with $\mathcal{H}^\pm$ purely-imaginary $2L\times 2L$ Hermitian matrices with indices defined as
\begin{multline}
	\mathcal{H}^\pm_{2m+l,2n+j} :=
	\delta_{m,n} 2 J h \sigma^y_{l,j} 
	+\\-
	(\delta_{m+1,n} \pm \delta_{m,0}\delta_{n,L-1})
	J (\sigma^y + i  \gamma \sigma^x)_{l,j} 
	+\\+ 
	(\delta_{m,n+1} \pm \delta_{m,L-1}\delta_{n,0})
	J (\sigma^y - i  \gamma \sigma^x)_{l,j}\,,
\end{multline}
for $m,n\in\{0,...,L-1\}$ and $l,j\in\{1,2\}$.
The fermionic theory described by $ H^+$ (resp.~$ H^-$) is called \textit{Ramond sector} (resp.~\textit{Neveu-Schwartz sector}) of the initial spin chain and has periodic (resp.~anti-periodic) boundary conditions.

The quadratic fermionic theories can be diagonalized by the Bogoliubov transformation 
\begin{equation}
	\label{eq:bogoliubov_tranformation_finite_size}
	\eta_k=\frac{1}{2\sqrt{L}} \sum_{j=0}^{L-1}
	e^{-i j k_\pm} (e^{i\theta_{k_\pm}/2}  a_{2j+1} - i e^{-i\theta_{k_\pm}/2}  a_{2j+2})
	,
\end{equation}
which allows one to recast the fermionic Hamiltonians in the form
\begin{equation}
	H^\pm=\sum_{k}\epsilon_{k_\pm}\left(\eta^\dag_k \eta_k -\frac{1}{2}\right)
 \label{e.Hpm}
\end{equation}
where the momenta $k_{\pm}$ are quantized depending on the sector as $k_\pm\in(-\pi,\pi]$ such that $e^{ i k_\pm L}=\pm1$, namely 
\begin{equation}
k_+ = 2n\pi/L~~~~~k_- = (2n-1)\pi/L
\end{equation}
with $n = -L/2+1,..., L/2$. Moreover
\begin{equation}
	\epsilon_k=2|J|\sqrt{(h- \cos(k))^2 + \gamma^2\sin^2(k)}
\end{equation}
is the dispersion relation, and $\theta_k\in(-\pi, \pi]$ is the Bogoliubov angle such that
\begin{equation}
\begin{cases}
	&\epsilon_k \sin(\theta_k)=-2J\gamma\sin(k)\ \\
 &\epsilon_k \cos(\theta_k)=2J(h-\cos(k))~.
\end{cases}
\end{equation}
The zero-frequency mode $\epsilon_k=0$ corresponds to the following condition on $\gamma$, $h$ and $k$: $\gamma\sin(k)=h-\cos(k)=0$. This condition leaves then the $\theta_k$ undetermined, and in this case we set $\theta_k=0$ by convention. 

In terms of the $\eta_k$ operators, the $\sigma_k^z$ operator takes the form  
\begin{multline}
 \sigma_k^z = \frac{2}{\sqrt{L}}  \sum_q c^\dagger_{q} c_{q+k} = \frac{2}{\sqrt{L}} \sum_q \big [ \cos\left(\tfrac{\theta_q+\theta_{q+k}}{2}\right) \eta^\dagger_q\eta_{q+k} 
\\
-\tfrac{i}{2} \eta_q \eta_{k-q} \sin\left(\tfrac{\theta_q-\theta_{k-q}}{2}\right) 
+\tfrac{i}{2} \eta^\dagger_{-k-q}\eta^\dagger_q
\sin\left(\tfrac{\theta_q-\theta_{-k-q}}{2}\right) 
\big ] ~,
\label{e.sigmakz}
 \end{multline}
namely the operator $\sigma_k^z$ creates (destroys) pairs of fermionic quasi-paricles $\eta$ with net momentum $-k$ ($+k$); or it creates particle-hole pairs with net momentum transfer of $-k$. 

\section{Time evolution}

\subsection{Gaussian states}
The non-equilibrium dynamics driven by the quadratic Hamiltonian Eq.~\eqref{e.Hpm} has the property of evolving Gaussian states into Gaussian states, namely states which are characterized by the fact of verifying Wick's theorem~\cite{Gaudin1960}. The consequence of the latter theorem is that any correlation function computed in the state can be expressed in terms of the two-point fermionic correlation function. 
It is then customary to define the correlation matrix
\begin{equation}
	\label{EQ: definition of the correlation matrix}
	\Gamma_{i,j}:= \delta_{i,j}-\braket{ a_i a_j}
	\,,
\end{equation}
where the expectation value is evaluated with respect to the Gaussian state of the system. Given Wick's theorem, the correlation matrix fully specifies the Gaussian state.

The time evolution of a Gaussian state is therefore entirely specified in terms of the evolution of its correlation matrix, obeying the equation~\cite{bocini_thesis}
\begin{equation}
	\Gamma(t)=
	e^{-i t\mathcal{H}} \Gamma(0) e^{i t\mathcal{H}}
	\,,
\end{equation}
where $\mathcal{H}$ can be either $\mathcal{H}^+$ or $\mathcal{H}^-$, depending on the sector under consideration.
Importantly, $\mathcal{H}$ can be diagonalized analytically as
\begin{equation}
	\mathcal{H}=U D U^\dag,
\end{equation}
where
\begin{equation}
	D_{j',j}=\delta_{j',j}\epsilon_{k_\pm(j)}\operatorname{sign}(L+1/2-j)
\end{equation}
for $j',j\in\{1,...,2L\}$, $k_\pm(j)=\frac{\pi (2j + \frac{1\mp 1}{2})}{L}$, and
\begin{align}
	U_{2m+1, j} =& \frac{1}{\sqrt{2}} e^{i m k_\pm(j) - i\theta_{k_\pm(j)}/2} 
	\\
	U_{2m+2, j} =& \frac{i}{\sqrt{2}} e^{i m k_\pm(j) + i\theta_{k_\pm(j)}/2} 
\end{align}
for $m\in\{0,...,L/2-1\}, j\in\{1,...,L\}$ and the other columns obtained by complex conjugation as $U_{a,j+L}=U^*_{a,j}$, $j\in\{1,...,L\}$.
Plugging in the expressions of the dispersion relation, the Bogoliubov angle and the correlation matrix describing the initial state, one can reconstruct the correlation matrix at all times.

\subsection{The initial states}

Any state with the spins aligned along the $z$-axis in the XY model considered above is Gaussian (regardless of the orientation of the spins), so, in particular, also the coherent spin state $\ket{{\rm CSS}_z}$, with all the spins pointing \textit{up}, is Gaussian. 
It belongs to the Neveu-Schwartz sector, since it is associated with a positive parity $O_{1,L}=1$; and its correlation matrix reads
\begin{align}
	\label{EQ: correlation matrix of spin states}
	& \Gamma^{(z)}_{2m+1,2n+1}=0=\Gamma^{(z)}_{2m+2,2n+2}, 
	\nonumber \\
	& \Gamma^{(z)}_{2m+1,2n+2}=
	- \Gamma^{(z)}_{2n+2,2m+1} =
	-i \delta_{m,n} ,
\end{align}
for all $m,n\in\{0,...,L-1\}$.

The state in which all the spins are aligned with the $x$ direction, $|{\rm CSS}_x\rangle$, on the other hand, is not Gaussian.
However, it can be easily written as the symmetric superposition of two Gaussian states. Indeed it can be viewed as belonging to the doubly-degenerate ground-state eigenspace of our Hamiltonian of interest with parameters $(J,\gamma,h)=(-1,1,0^+)$, namely the classical Ising limit (in an infinitesimal transverse field). The two degenerate ground states in this limit are two Schrödinger's cat states 
\begin{equation}
|{\rm GS}^{\pm}\rangle = \frac{1}{\sqrt{2}} \left ( |{\rm CSS}_x\rangle \pm |{\rm CSS}_{-x}\rangle \right )~
\end{equation}
 which represent the ground states of the Ramond and Neveu-Schwartz sectors respectively.
One can easily write $|{\rm CSS}_x\rangle$ in terms of these states 
\begin{equation}
	\ket{{\rm CSS}_x} = \frac{1}{\sqrt 2} \left ( \ket{{\rm GS}^+} + \ket{{\rm GS}^-} \right )~.
\end{equation}
where $\ket{{\rm GS}^+}$ and $\ket{{\rm GS}^-}$ are the (Gaussian) ground states of the Ramond and Neveu-Schwartz sectors respectively.

The correlation matrices for the $\ket{{\rm GS}^\pm}$ states are 
\begin{align}
	&\Gamma^{(\pm)}_{2m+1,2n+1}=0=\Gamma^{(\pm)}_{2m+2,2n+2}, 
	\\
	&\Gamma^{(\pm)}_{2m+1,2n+2}=-\Gamma^{(\pm)}_{2n+2,2m+1}
	=
	i\delta_{m,n+1}\pm i\delta_{m,0},\delta_{n,L-1}, \nonumber
\end{align}
where the $\pm$ signs are associated with the  $\ket{{\rm GS}^\pm}$ states.

\subsection{From fermions to spins}

As already discussed above, the time evolution of the Gaussian states $|{\rm CSS}_z\rangle$ and  $\ket{{\rm GS}^\pm}$ states can be easily computed in terms of the evolution of their correlation matrices. 
In the case of $|{\rm CSS}_z\rangle$ as initial state, the evolution is driven by the Hamiltonian $H^{-}$, and this  allows one to calculate the time evolution of spin observables in terms of the evolved $\Gamma$ matrix, as we will detail below. 

When initializing the dynamics in the $|{\rm CSS}_x\rangle$, instead, the evolved state is the symmetric superposition of the evolution of $\ket{{\rm GS}^+}$ using $H^{+}$ and 
of $\ket{{\rm GS}^-}$ using $H^{-}$.  

Throughout this work we restrict ourselves to observables $A$ that commute with the parity operator $\mathcal{O}_{1,L}$, namely which cannot admix the two sectors of the fermionic theory. Hence the evolution of these observables simply writes as the average of the evolution of the same observables in the two fermionic sectors: 
\begin{align}
 \langle A \rangle(t) & = \braket{{\rm CSS}_x|e^{iHt} ~A~ e^{-iHt}|{\rm CSS}_x}  \nonumber \\
& = \frac{1}{2} \sum_{\alpha = \pm} \braket{{\rm GS}^\alpha|e^{iH^\alpha t}~ A ~e^{-iH^\alpha t}|{\rm GS}^\alpha } 
\end{align}
This crucial property allows us to conveniently perform all computations within the Gaussian-state formalism even when the initial state is not Gaussian. 

Below we provide the expressions of the observables studied in the main text in terms of the correlation matrix for Gaussian states, making use of Wick's theorem:
\begin{align}
	C_F(\ell, t) & \equiv
	\frac{1}{2} \braket{
		\left ( \sigma^y_1 \sigma^y_{\ell+1} - \sigma^x_1 \sigma^x_{\ell+1} \right ) 
		\mathcal{O}_{2,\ell}}_\alpha
 \nonumber \\ & =
	\frac{i}{2}	
	\left(\Gamma^{(\alpha)}_{2\ell+1,2}(t)
	+ \Gamma^{(\alpha)}_{2\ell+2,1}(t)\right),
\end{align}
\begin{align}
	& \braket{\sigma^z_m\sigma^z_n}_\alpha(t) = \delta_{m,n}+
	\Gamma^{(\alpha)}_{2m-1,2n}(t)\Gamma^{(\alpha)}_{2m,2n-1}(t)
	\\
	& -\Gamma^{(\alpha)}_{2m-1,2n-1}(t)\Gamma^{(\alpha)}_{2m,2n}(t)
	+\Gamma^{(\alpha)}_{2m-1,2m}(t)\Gamma^{(\alpha)}_{2n-1,2n}(t), \nonumber 
\end{align}
\begin{equation}
	\braket{\sigma^z_m}_\alpha(t)= 
	i\Gamma^{(\alpha)}_{2m-1,2m}(t),
\end{equation}
\begin{equation}
	\braket{\sigma^x_m\sigma^x_n}_\alpha(t) =
	\operatorname{Pf}(iB^{(mn,\alpha)}),
\end{equation}
where $\alpha =$ +, - and $z$, corresponding to one of the Gaussian states described above, and $\langle ... \rangle_{\alpha}$ is the average over the state starting from the Gaussian state labeled by $\alpha$;  $\operatorname{Pf}$ is the Pfaffian and, assuming $m<n$, $B^{(mn,\alpha)}$ is the sub-matrix of $\Gamma^{\alpha}$ defined as $B^{mn,\alpha}_{l,j}=\Gamma^{(\alpha)}_{2m-1+l,2m-1+j}$, for $l,j\in\{1,...,2(n-m)\}$.
These expressions are the final ingredient needed to reproduce the results shown in the main text.

\subsection{Invariant sub-spaces}

The arguments discussed above apply to any $\Gamma$, but in practice, translational invariance (of both the initial state and the Hamiltonian) allows to simplify the computation of the fermionic quench spectroscopy even further, following the formalism of Ref.~\cite{Fagotti_2020,Bocini_2023}.
Essentially, any correlation matrix describing a homogeneous state can be split into two independent contributions as
\begin{multline}
    \Gamma_{2m+j,2n+l}=\Gamma_{2m+j,2n+l}^\rho + \Gamma^\psi_{2m+j,2n+l}
    \\\equiv
    \frac{1}{L}
    \sum_k\left(\hat\Gamma_{jl}^\rho(k_\pm) + \hat\Gamma^\psi_{jl}(k_\pm)\right)e^{i(m-n)k_\pm},
\end{multline}
with $m,n\in\{0,...,L-1\}$ and $j,l\in\{1,2\}$.
Here $\hat\Gamma^\rho(k)$ and $\hat\Gamma^\psi(k)$ are $2\times2$ matrices defined as
\begin{equation}
    \hat\Gamma^\rho(k)=e^{-i\frac{\theta_k}{2}\sigma^z}[(4\pi\tfrac{\rho_k+\rho_{-k}}{2}-1)\sigma^y+4\pi\tfrac{\rho_k-\rho_{-k}}{2} \mathbb{I}]e^{i\frac{\theta_k}{2}\sigma^z},
\end{equation}
\begin{equation}
    \hat\Gamma^\psi(k)=e^{-i\frac{\theta_k}{2}\sigma^z}[\operatorname{Re}(\psi_k)\sigma^z -  \operatorname{Im}(\psi_k)\sigma^x]e^{-i\frac{\theta_{-k}}{2}\sigma^z},
\end{equation}
where $\rho_k :=\frac{\braket{\eta_k^\dagger\eta_k}}{2\pi}$, $\psi_k:=-2\braket{\eta_{-k}\eta_k}$ respectively (by definition $\psi_\pi=0$).
Such a decomposition is particularly convenient because the full time dependence of the state is encoded in the scalar function $\psi_k$, while $\Gamma^\rho$ is time independent.
This implies that $\Gamma^\rho$ does not contribute to $Q_F(k,\omega\neq0)$ and therefore, as far as fermionic quench spectroscopy is concerned, we can restrict ourselves to study only the $\Gamma^\psi$ part of any initial correlation matrix.
A clear signature of this is the result in the infinite-size limit
\begin{multline}
\label{eq:lim_Q_F}
    Q_F(k,\omega\neq0)=\\=-\pi\cos\left(\tfrac{\theta_k-\theta_{-k}}{2}\right) \operatorname{Im}(\psi_k) (\delta(\omega+2\epsilon_k)
    + \delta(\omega-2\epsilon_k))
\end{multline}
where $\psi_k$ describes the initial correlation matrix (see below for more details about the infinite-size limit).
In this limit, it can be shown from the definitions above that $\operatorname{Im}(\psi_k)=-\sin \left(k-\frac{\theta_k-\theta_{-k}}{2}\right )$ if the initial state is $\ket{{\rm CSS}_x}$ and $\operatorname{Im}(\psi_k)=-\sin(\frac{\theta_k-\theta_{-k}}{2})$ if the initial state is $\ket{{\rm CSS}_z}$. 

Incidentally, we notice that the prediction for the dynamics of the Ising chain starting from the $\ket{{\rm CSS}_z}$ state is a $Q_F(k,\omega\neq0)$ whose sign is modulated by the factor $(\cos k-h)\sin k$; this explains the change in sign shown in Fig.~2 of the main text for $h<1$.

\section{Definition of the Fourier transforms}

The explicit definitions of the Fourier transforms (FTs) of the fermionic correlations that we use are as follows:
\begin{equation}
    S_F(k,t)=-2\sum_{\ell=1}^{L/2-1}\sin(k\ell)C_F(\ell,t),
\end{equation}
where the FT is taken with the $\sin$ function because of the antisymmetric nature of the fermionic correlation function; and 
\begin{equation}
    Q_F(k,\omega)=2\sum_{j=1}^{N-1}\cos(\omega j\delta_t) S_F(k,j \delta_t)\,,
\end{equation}
for $\omega =\frac{2\pi}{2N-1}\frac{\nu}{\delta_t}$, $\nu\in\{-N+1,...,N-1\}$, $N=T/\delta_t$, where $T$ is the total evolution time and $\delta_t$ is the discrete time step.


\begin{figure}[ht!]
    \centering
    \includegraphics[width=0.95\linewidth]{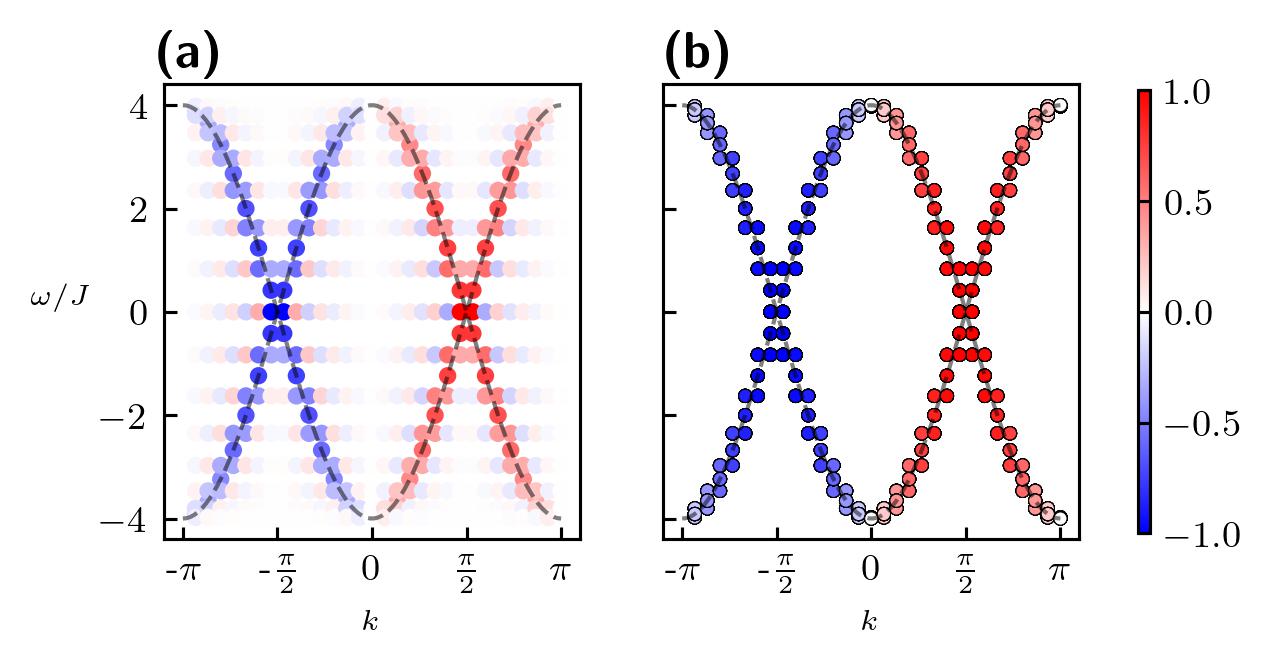}
    \caption{(a) Normalized fermionic QSF $Q_F(k,\omega)$ for the XX chain with $L=30$ in the infinite-time limit. (b) 
    Values of $\epsilon_{k_\pm}$, with associated spectral weights given by $\sin(k)$.
    In both panels, spectral weights have been normalized to the maximum absolute value. }
    \label{fig:fig:infinite_time}
\end{figure}

\begin{figure*}[ht!]
    \centering
    \includegraphics[width=0.8\textwidth]{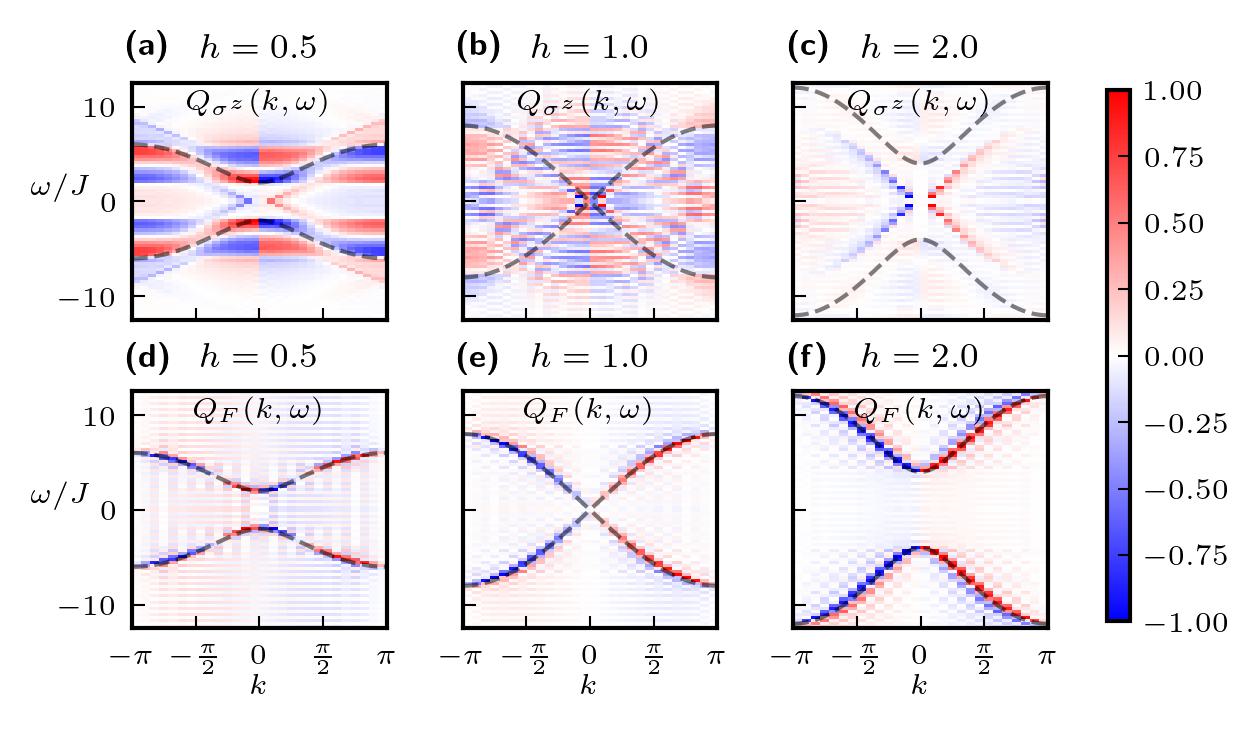}
    \caption{\emph{Quench spectral functions for the quantum Ising chain starting from the $|{\rm CSS}_x\rangle$ state.} 
    (a-c) Spin QSF $Q_{\sigma^z}(k,\omega)$ for three values of the transverse field $h = 0.5, 1$ and 2; (d-f) Fermionic QSF $Q_F(k,\omega)$ for the same field values. Dashed lines indicate the dispersion relation of fermionic quasiparticles. All conventions for the Fourier transform and the normalization of the spectral weights are as in the figures of the main text.}
    \label{fig:ZZ_CSSx}
\end{figure*}

\section{Infinite-size limit}

Looking at the infinite-size limit of $Q_F(k,\omega)$ makes the spectral content of the correlation clearer.
In practice, we take first the limit $L\rightarrow\infty$, then we also consider $T\rightarrow\infty$ and $dt\rightarrow0$ to have perfect resolution in frequencies. The spectral function can now be expressed as
\begin{equation}
    Q_F(k,\omega) \rightarrow -\int_{-\infty}^{+\infty} dt
    \cos(\omega t)\sum_{\ell=1}^{\infty}2\sin(\ell k)C_F(\ell, t),
\end{equation}
and from here one obtains results such as the one expressed in Eq.~\eqref{eq:lim_Q_F}.

When comparing finite results with such a limit, one should keep in mind that the infinite-size limit in this case is taken \emph{before} sending the evolution time to infinity.
In this way, because of Lieb-Robinson bounds on the propagation of information~\cite{Lieb.Robinson1972}, the dynamics of correlations never probe the boundaries of the system for a finite time. 
Importantly, the two limits (infinite size and infinite time) do not commute.
In finite systems, this means that we can consider time evolution only up to times that are smaller than the system size.

We note however that the case in which one evolves the system for a time comparable to the system size is also interesting.
In this case, the system's dynamics are affected by finite-size effects. In practice, this allows one to distinguish the contributions of the two fermionic sectors, as shown in Fig.~\ref{fig:fig:infinite_time}(a), where we take the limit $T\rightarrow\infty$ keeping $L=30$ in the XX chain.
We see that the contribution of the $H^+$ fermionic sector, at wavevectors $k = 2\pi n/L$, falls on top of the infinite-size prediction for the frequencies of the fermionic quasiparticles, namely $\epsilon_k = \pm2|J|\cos(2 n\pi /L)$; while the contribution from the $H^-$ sector gives a contribution which is peaked around the infinite-size frequencies, but it has a finite width. Indeed the corresponding frequencies are of the kind $\epsilon_k = \pm2|J|\cos((2n-1)\pi/L)$, whose associated wavevectors miss by a factor $O(1/L)$  the ones admitted on a ring with  periodic boundary conditions. Fig.~\ref{fig:fig:infinite_time}(b) shows that the dominant frequency components are recovered as $\omega_{k = 2\pi n/L} = \pm 4|J|\cos(2 n\pi /L), \pm 4|J|\cos((2n\pm 1)\pi/L)$ upon varying the integer index $n$.


\section{Quench spectroscopy of the Ising model starting from the ${\rm CSS}_x$ state}

To complement the study of the quench spectroscopy for the Ising model reported in the main text, we consider the case of the $|{\rm CSS}_x\rangle$ state as initial state of the quench dynamics. This state is a low-energy state for the ordered phase of the system, namely for $h<1$. 
Fig.~\ref{fig:ZZ_CSSx} shows both the spin quench spectral function (QSF) 
$Q_{\sigma^z}(k,\omega)$ as well as the fermionic one $Q_F(k,\omega)$ for various field values. 
While nonlocal quench spectroscopy based on the fermonic QSF reconstructs perfectly the quasi-particle dispersion relation in this case as well, the conventional quench spectroscopy based on the spin QSF appears to fail rather clearly to deliver the correct picture for the excitation spectrum. In particular, the spin QSF has a very complex frequency structure, with little relationship to the fermionic excitation spectrum, and, away from the critical point $h=1$, showing frequencies which lie systematically inside the gap of the fermionic spectrum. This is actually not too surprising when considering that the spin operator $\sigma_k^z$ creates particle-hole pairs, as shown in Eq.~\eqref{e.sigmakz}. The initial state of the evolution is not sensitive to the gapped spectrum, namely it is the superposition of Fock states of the $\eta$ quasiparticles which live both above and below the gap. This means that the Hamiltonian eigenstates $|m\rangle, |n\rangle$ (with  $\langle m | {\rm CSS}_x\rangle, \langle n | {\rm CSS}_x\rangle \neq 0$) connected by the creation of two particle-hole pairs, and contributing to the the spin QSF, can be arbitrarily close in energy, as authorized by the difference $\epsilon_{q_1-k} + \epsilon_{q_2+k} - \epsilon_{q_1} - \epsilon_{q_2}$ for any pair of momenta $q_1, q_2$ with a finite quasi-particle population in the initial state.  

Therefore conventional quench spectroscopy in the case of the Ising model can fail clearly in reconstructing the low-energy excitations of the system defined above the ground state. This represents a rather compelling evidence of the special nature of the excitations in this system, and of the relevance of nonlocal quench spectroscopy based on fermionic correlation functions.


  \bibliography{references}%

\begin{thebibliography}{44}
\expandafter\ifx\csname natexlab\endcsname\relax\def\natexlab#1{#1}\fi
\expandafter\ifx\csname bibnamefont\endcsname\relax
  \def\bibnamefont#1{#1}\fi
\expandafter\ifx\csname bibfnamefont\endcsname\relax
  \def\bibfnamefont#1{#1}\fi
\expandafter\ifx\csname citenamefont\endcsname\relax
  \def\citenamefont#1{#1}\fi
\expandafter\ifx\csname url\endcsname\relax
  \def\url#1{\texttt{#1}}\fi
\expandafter\ifx\csname urlprefix\endcsname\relax\def\urlprefix{URL }\fi
\providecommand{\bibinfo}[2]{#2}
\providecommand{\eprint}[2][]{\url{#2}}

\bibitem[{\citenamefont{Sachdev}(2023)}]{Sachdev-book}
\bibinfo{author}{\bibfnamefont{S.}~\bibnamefont{Sachdev}},
  \emph{\bibinfo{title}{Quantum Phases of Matter}}
  (\bibinfo{publisher}{Cambridge}, \bibinfo{year}{2023}).

\bibitem[{\citenamefont{Boothroyd}(2020)}]{Boothroyd-book}
\bibinfo{author}{\bibfnamefont{A.~T.} \bibnamefont{Boothroyd}},
  \emph{\bibinfo{title}{Principles of Neutron Scattering from Condensed
  Matter}} (\bibinfo{publisher}{UOP Oxford}, \bibinfo{year}{2020}).

\bibitem[{\citenamefont{Blundell}(2001)}]{Blundell2001}
\bibinfo{author}{\bibfnamefont{S.}~\bibnamefont{Blundell}},
  \emph{\bibinfo{title}{Magnetism in Condensed Matter}}
  (\bibinfo{publisher}{UOP Oxford}, \bibinfo{year}{2001}).

\bibitem[{\citenamefont{Slichter}(2008)}]{Slichter-book}
\bibinfo{author}{\bibfnamefont{C.~P.} \bibnamefont{Slichter}},
  \emph{\bibinfo{title}{Principles of Magnetic Resonance}}
  (\bibinfo{publisher}{Springer}, \bibinfo{year}{2008}).

\bibitem[{\citenamefont{Blundell et~al.}(2021)\citenamefont{Blundell, De~Renzi,
  Lancaster, and Pratt}}]{Muon-book}
\bibinfo{author}{\bibfnamefont{S.~J.} \bibnamefont{Blundell}},
  \bibinfo{author}{\bibfnamefont{R.}~\bibnamefont{De~Renzi}},
  \bibinfo{author}{\bibfnamefont{T.}~\bibnamefont{Lancaster}},
  \bibnamefont{and} \bibinfo{author}{\bibfnamefont{F.~L.} \bibnamefont{Pratt}},
  \emph{\bibinfo{title}{Muon Spectroscopy -- An Introduction}}
  (\bibinfo{publisher}{UOP Oxford}, \bibinfo{year}{2021}).

\bibitem[{\citenamefont{Savary and Balents}(2016)}]{Savary2017}
\bibinfo{author}{\bibfnamefont{L.}~\bibnamefont{Savary}} \bibnamefont{and}
  \bibinfo{author}{\bibfnamefont{L.}~\bibnamefont{Balents}},
  \bibinfo{journal}{Reports on Progress in Physics}
  \textbf{\bibinfo{volume}{80}}, \bibinfo{pages}{016502}
  (\bibinfo{year}{2016}),
  \urlprefix\url{https://dx.doi.org/10.1088/0034-4885/80/1/016502}.

\bibitem[{\citenamefont{Lake et~al.}(2005)\citenamefont{Lake, Tennant, Frost,
  and Nagler}}]{Lake2005}
\bibinfo{author}{\bibfnamefont{B.}~\bibnamefont{Lake}},
  \bibinfo{author}{\bibfnamefont{D.~A.} \bibnamefont{Tennant}},
  \bibinfo{author}{\bibfnamefont{C.~D.} \bibnamefont{Frost}}, \bibnamefont{and}
  \bibinfo{author}{\bibfnamefont{S.~E.} \bibnamefont{Nagler}},
  \bibinfo{journal}{Nature Materials} \textbf{\bibinfo{volume}{4}},
  \bibinfo{pages}{329} (\bibinfo{year}{2005}), ISSN \bibinfo{issn}{1476-4660},
  \urlprefix\url{https://doi.org/10.1038/nmat1327}.

\bibitem[{\citenamefont{Coldea et~al.}(2010)\citenamefont{Coldea, Tennant,
  Wheeler, Wawrzynska, Prabhakaran, Telling, Habicht, Smeibidl, and
  Kiefer}}]{Coldea2011}
\bibinfo{author}{\bibfnamefont{R.}~\bibnamefont{Coldea}},
  \bibinfo{author}{\bibfnamefont{D.~A.} \bibnamefont{Tennant}},
  \bibinfo{author}{\bibfnamefont{E.~M.} \bibnamefont{Wheeler}},
  \bibinfo{author}{\bibfnamefont{E.}~\bibnamefont{Wawrzynska}},
  \bibinfo{author}{\bibfnamefont{D.}~\bibnamefont{Prabhakaran}},
  \bibinfo{author}{\bibfnamefont{M.}~\bibnamefont{Telling}},
  \bibinfo{author}{\bibfnamefont{K.}~\bibnamefont{Habicht}},
  \bibinfo{author}{\bibfnamefont{P.}~\bibnamefont{Smeibidl}}, \bibnamefont{and}
  \bibinfo{author}{\bibfnamefont{K.}~\bibnamefont{Kiefer}},
  \bibinfo{journal}{Science} \textbf{\bibinfo{volume}{327}},
  \bibinfo{pages}{177} (\bibinfo{year}{2010}),
  \eprint{https://www.science.org/doi/pdf/10.1126/science.1180085},
  \urlprefix\url{https://www.science.org/doi/abs/10.1126/science.1180085}.

\bibitem[{\citenamefont{Han et~al.}(2012)\citenamefont{Han, Helton, Chu,
  Nocera, Rodriguez-Rivera, Broholm, and Lee}}]{Han2012}
\bibinfo{author}{\bibfnamefont{T.-H.} \bibnamefont{Han}},
  \bibinfo{author}{\bibfnamefont{J.~S.} \bibnamefont{Helton}},
  \bibinfo{author}{\bibfnamefont{S.}~\bibnamefont{Chu}},
  \bibinfo{author}{\bibfnamefont{D.~G.} \bibnamefont{Nocera}},
  \bibinfo{author}{\bibfnamefont{J.~A.} \bibnamefont{Rodriguez-Rivera}},
  \bibinfo{author}{\bibfnamefont{C.}~\bibnamefont{Broholm}}, \bibnamefont{and}
  \bibinfo{author}{\bibfnamefont{Y.~S.} \bibnamefont{Lee}},
  \bibinfo{journal}{Nature} \textbf{\bibinfo{volume}{492}},
  \bibinfo{pages}{406} (\bibinfo{year}{2012}), ISSN \bibinfo{issn}{1476-4687},
  \urlprefix\url{https://doi.org/10.1038/nature11659}.

\bibitem[{\citenamefont{Laurell et~al.}(2021)\citenamefont{Laurell, Scheie,
  Mukherjee, Koza, Enderle, Tylczynski, Okamoto, Coldea, Tennant, and
  Alvarez}}]{Laurell2021}
\bibinfo{author}{\bibfnamefont{P.}~\bibnamefont{Laurell}},
  \bibinfo{author}{\bibfnamefont{A.}~\bibnamefont{Scheie}},
  \bibinfo{author}{\bibfnamefont{C.~J.} \bibnamefont{Mukherjee}},
  \bibinfo{author}{\bibfnamefont{M.~M.} \bibnamefont{Koza}},
  \bibinfo{author}{\bibfnamefont{M.}~\bibnamefont{Enderle}},
  \bibinfo{author}{\bibfnamefont{Z.}~\bibnamefont{Tylczynski}},
  \bibinfo{author}{\bibfnamefont{S.}~\bibnamefont{Okamoto}},
  \bibinfo{author}{\bibfnamefont{R.}~\bibnamefont{Coldea}},
  \bibinfo{author}{\bibfnamefont{D.~A.} \bibnamefont{Tennant}},
  \bibnamefont{and} \bibinfo{author}{\bibfnamefont{G.}~\bibnamefont{Alvarez}},
  \bibinfo{journal}{Phys. Rev. Lett.} \textbf{\bibinfo{volume}{127}},
  \bibinfo{pages}{037201} (\bibinfo{year}{2021}),
  \urlprefix\url{https://link.aps.org/doi/10.1103/PhysRevLett.127.037201}.

\bibitem[{\citenamefont{Browaeys and Lahaye}(2020)}]{Browaeys2020}
\bibinfo{author}{\bibfnamefont{A.}~\bibnamefont{Browaeys}} \bibnamefont{and}
  \bibinfo{author}{\bibfnamefont{T.}~\bibnamefont{Lahaye}},
  \bibinfo{journal}{Nature Physics} \textbf{\bibinfo{volume}{16}},
  \bibinfo{pages}{132} (\bibinfo{year}{2020}), ISSN \bibinfo{issn}{1745-2481},
  \urlprefix\url{https://doi.org/10.1038/s41567-019-0733-z}.

\bibitem[{\citenamefont{Monroe et~al.}(2021)\citenamefont{Monroe, Campbell,
  Duan, Gong, Gorshkov, Hess, Islam, Kim, Linke, Pagano et~al.}}]{Monroe_2021}
\bibinfo{author}{\bibfnamefont{C.}~\bibnamefont{Monroe}},
  \bibinfo{author}{\bibfnamefont{W.~C.} \bibnamefont{Campbell}},
  \bibinfo{author}{\bibfnamefont{L.-M.} \bibnamefont{Duan}},
  \bibinfo{author}{\bibfnamefont{Z.-X.} \bibnamefont{Gong}},
  \bibinfo{author}{\bibfnamefont{A.~V.} \bibnamefont{Gorshkov}},
  \bibinfo{author}{\bibfnamefont{P.~W.} \bibnamefont{Hess}},
  \bibinfo{author}{\bibfnamefont{R.}~\bibnamefont{Islam}},
  \bibinfo{author}{\bibfnamefont{K.}~\bibnamefont{Kim}},
  \bibinfo{author}{\bibfnamefont{N.~M.} \bibnamefont{Linke}},
  \bibinfo{author}{\bibfnamefont{G.}~\bibnamefont{Pagano}},
  \bibnamefont{et~al.}, \bibinfo{journal}{Reviews of Modern Physics}
  \textbf{\bibinfo{volume}{93}} (\bibinfo{year}{2021}), ISSN
  \bibinfo{issn}{1539-0756},
  \urlprefix\url{http://dx.doi.org/10.1103/RevModPhys.93.025001}.

\bibitem[{\citenamefont{Garc\'ia~Ripoll}(2022)}]{Garcia-Ripoll-book}
\bibinfo{author}{\bibfnamefont{J.~J.} \bibnamefont{Garc\'ia~Ripoll}},
  \emph{\bibinfo{title}{Quantum Information and Quantum Optics with
  Superconducting Circuits}} (\bibinfo{publisher}{Cambridge},
  \bibinfo{year}{2022}).

\bibitem[{\citenamefont{Menu and Roscilde}(2018)}]{MenuR2018}
\bibinfo{author}{\bibfnamefont{R.}~\bibnamefont{Menu}} \bibnamefont{and}
  \bibinfo{author}{\bibfnamefont{T.}~\bibnamefont{Roscilde}},
  \bibinfo{journal}{Phys. Rev. B} \textbf{\bibinfo{volume}{98}},
  \bibinfo{pages}{205145} (\bibinfo{year}{2018}),
  \urlprefix\url{https://link.aps.org/doi/10.1103/PhysRevB.98.205145}.

\bibitem[{\citenamefont{Villa et~al.}(2019)\citenamefont{Villa, Despres, and
  Sanchez-Palencia}}]{Villa2019}
\bibinfo{author}{\bibfnamefont{L.}~\bibnamefont{Villa}},
  \bibinfo{author}{\bibfnamefont{J.}~\bibnamefont{Despres}}, \bibnamefont{and}
  \bibinfo{author}{\bibfnamefont{L.}~\bibnamefont{Sanchez-Palencia}},
  \bibinfo{journal}{Phys. Rev. A} \textbf{\bibinfo{volume}{100}},
  \bibinfo{pages}{063632} (\bibinfo{year}{2019}),
  \urlprefix\url{https://link.aps.org/doi/10.1103/PhysRevA.100.063632}.

\bibitem[{\citenamefont{Villa et~al.}(2020)\citenamefont{Villa, Despres,
  Thomson, and Sanchez-Palencia}}]{Villa2020}
\bibinfo{author}{\bibfnamefont{L.}~\bibnamefont{Villa}},
  \bibinfo{author}{\bibfnamefont{J.}~\bibnamefont{Despres}},
  \bibinfo{author}{\bibfnamefont{S.~J.} \bibnamefont{Thomson}},
  \bibnamefont{and}
  \bibinfo{author}{\bibfnamefont{L.}~\bibnamefont{Sanchez-Palencia}},
  \bibinfo{journal}{Phys. Rev. A} \textbf{\bibinfo{volume}{102}},
  \bibinfo{pages}{033337} (\bibinfo{year}{2020}),
  \urlprefix\url{https://link.aps.org/doi/10.1103/PhysRevA.102.033337}.

\bibitem[{\citenamefont{Menu and Roscilde}(2023)}]{Menu2023}
\bibinfo{author}{\bibfnamefont{R.}~\bibnamefont{Menu}} \bibnamefont{and}
  \bibinfo{author}{\bibfnamefont{T.}~\bibnamefont{Roscilde}},
  \bibinfo{journal}{SciPost Phys.} \textbf{\bibinfo{volume}{14}},
  \bibinfo{pages}{151} (\bibinfo{year}{2023}),
  \urlprefix\url{https://scipost.org/10.21468/SciPostPhys.14.6.151}.

\bibitem[{\citenamefont{Chen et~al.}(2023)\citenamefont{Chen, Emperauger,
  Bornet, Caleca, G\'ely, Bintz, Chatterjee, Liu, Barredo, Yao
  et~al.}}]{chen2023spectroscopy}
\bibinfo{author}{\bibfnamefont{C.}~\bibnamefont{Chen}},
  \bibinfo{author}{\bibfnamefont{G.}~\bibnamefont{Emperauger}},
  \bibinfo{author}{\bibfnamefont{G.}~\bibnamefont{Bornet}},
  \bibinfo{author}{\bibfnamefont{F.}~\bibnamefont{Caleca}},
  \bibinfo{author}{\bibfnamefont{B.}~\bibnamefont{G\'ely}},
  \bibinfo{author}{\bibfnamefont{M.}~\bibnamefont{Bintz}},
  \bibinfo{author}{\bibfnamefont{S.}~\bibnamefont{Chatterjee}},
  \bibinfo{author}{\bibfnamefont{V.}~\bibnamefont{Liu}},
  \bibinfo{author}{\bibfnamefont{D.}~\bibnamefont{Barredo}},
  \bibinfo{author}{\bibfnamefont{N.~Y.} \bibnamefont{Yao}},
  \bibnamefont{et~al.}, \emph{\bibinfo{title}{Spectroscopy of elementary
  excitations from quench dynamics in a dipolar xy rydberg simulator}}
  (\bibinfo{year}{2023}), \eprint{2311.11726}.

\bibitem[{\citenamefont{Jordan and Wigner}(1928)}]{Jordan:1928wi}
\bibinfo{author}{\bibfnamefont{P.}~\bibnamefont{Jordan}} \bibnamefont{and}
  \bibinfo{author}{\bibfnamefont{E.~P.} \bibnamefont{Wigner}},
  \bibinfo{journal}{Z. Phys.} \textbf{\bibinfo{volume}{47}},
  \bibinfo{pages}{631} (\bibinfo{year}{1928}).

\bibitem[{\citenamefont{Lieb et~al.}(1961)\citenamefont{Lieb, Schultz, and
  Mattis}}]{LSM1961}
\bibinfo{author}{\bibfnamefont{E.}~\bibnamefont{Lieb}},
  \bibinfo{author}{\bibfnamefont{T.}~\bibnamefont{Schultz}}, \bibnamefont{and}
  \bibinfo{author}{\bibfnamefont{D.}~\bibnamefont{Mattis}},
  \bibinfo{journal}{Annals of Physics} \textbf{\bibinfo{volume}{16}},
  \bibinfo{pages}{407} (\bibinfo{year}{1961}), ISSN \bibinfo{issn}{0003-4916},
  \urlprefix\url{https://www.sciencedirect.com/science/article/pii/0003491661901154}.

\bibitem[{\citenamefont{Franchini}(2017)}]{Franchini_2017}
\bibinfo{author}{\bibfnamefont{F.}~\bibnamefont{Franchini}},
  \emph{\bibinfo{title}{An Introduction to Integrable Techniques for
  One-Dimensional Quantum Systems}} (\bibinfo{publisher}{Springer International
  Publishing}, \bibinfo{year}{2017}), ISBN \bibinfo{isbn}{9783319484877},
  \urlprefix\url{http://dx.doi.org/10.1007/978-3-319-48487-7}.

\bibitem[{SM()}]{SM}
\bibinfo{note}{See {S}upplemental {M}aterial ({SM}) for details about: 1), 2)
  ... .}

\bibitem[{\citenamefont{Simon et~al.}(2011)\citenamefont{Simon, Bakr, Ma, Tai,
  Preiss, and Greiner}}]{Simon2011}
\bibinfo{author}{\bibfnamefont{J.}~\bibnamefont{Simon}},
  \bibinfo{author}{\bibfnamefont{W.~S.} \bibnamefont{Bakr}},
  \bibinfo{author}{\bibfnamefont{R.}~\bibnamefont{Ma}},
  \bibinfo{author}{\bibfnamefont{M.~E.} \bibnamefont{Tai}},
  \bibinfo{author}{\bibfnamefont{P.~M.} \bibnamefont{Preiss}},
  \bibnamefont{and} \bibinfo{author}{\bibfnamefont{M.}~\bibnamefont{Greiner}},
  \bibinfo{journal}{Nature} \textbf{\bibinfo{volume}{472}},
  \bibinfo{pages}{307} (\bibinfo{year}{2011}), ISSN \bibinfo{issn}{1476-4687},
  \urlprefix\url{https://doi.org/10.1038/nature09994}.

\bibitem[{\citenamefont{Labuhn et~al.}(2016)\citenamefont{Labuhn, Barredo,
  Ravets, de~L{\'e}s{\'e}leuc, Macr{\`i}, Lahaye, and Browaeys}}]{Labuhn2016}
\bibinfo{author}{\bibfnamefont{H.}~\bibnamefont{Labuhn}},
  \bibinfo{author}{\bibfnamefont{D.}~\bibnamefont{Barredo}},
  \bibinfo{author}{\bibfnamefont{S.}~\bibnamefont{Ravets}},
  \bibinfo{author}{\bibfnamefont{S.}~\bibnamefont{de~L{\'e}s{\'e}leuc}},
  \bibinfo{author}{\bibfnamefont{T.}~\bibnamefont{Macr{\`i}}},
  \bibinfo{author}{\bibfnamefont{T.}~\bibnamefont{Lahaye}}, \bibnamefont{and}
  \bibinfo{author}{\bibfnamefont{A.}~\bibnamefont{Browaeys}},
  \bibinfo{journal}{Nature} \textbf{\bibinfo{volume}{534}},
  \bibinfo{pages}{667} (\bibinfo{year}{2016}), ISSN \bibinfo{issn}{1476-4687},
  \urlprefix\url{https://doi.org/10.1038/nature18274}.

\bibitem[{\citenamefont{Bernien et~al.}(2017)\citenamefont{Bernien, Schwartz,
  Keesling, Levine, Omran, Pichler, Choi, Zibrov, Endres, Greiner
  et~al.}}]{Bernien2017}
\bibinfo{author}{\bibfnamefont{H.}~\bibnamefont{Bernien}},
  \bibinfo{author}{\bibfnamefont{S.}~\bibnamefont{Schwartz}},
  \bibinfo{author}{\bibfnamefont{A.}~\bibnamefont{Keesling}},
  \bibinfo{author}{\bibfnamefont{H.}~\bibnamefont{Levine}},
  \bibinfo{author}{\bibfnamefont{A.}~\bibnamefont{Omran}},
  \bibinfo{author}{\bibfnamefont{H.}~\bibnamefont{Pichler}},
  \bibinfo{author}{\bibfnamefont{S.}~\bibnamefont{Choi}},
  \bibinfo{author}{\bibfnamefont{A.~S.} \bibnamefont{Zibrov}},
  \bibinfo{author}{\bibfnamefont{M.}~\bibnamefont{Endres}},
  \bibinfo{author}{\bibfnamefont{M.}~\bibnamefont{Greiner}},
  \bibnamefont{et~al.}, \bibinfo{journal}{Nature}
  \textbf{\bibinfo{volume}{551}}, \bibinfo{pages}{579} (\bibinfo{year}{2017}),
  ISSN \bibinfo{issn}{1476-4687},
  \urlprefix\url{https://doi.org/10.1038/nature24622}.

\bibitem[{\citenamefont{King et~al.}(2022)\citenamefont{King, Suzuki, Raymond,
  Zucca, Lanting, Altomare, Berkley, Ejtemaee, Hoskinson, Huang
  et~al.}}]{King2022}
\bibinfo{author}{\bibfnamefont{A.~D.} \bibnamefont{King}},
  \bibinfo{author}{\bibfnamefont{S.}~\bibnamefont{Suzuki}},
  \bibinfo{author}{\bibfnamefont{J.}~\bibnamefont{Raymond}},
  \bibinfo{author}{\bibfnamefont{A.}~\bibnamefont{Zucca}},
  \bibinfo{author}{\bibfnamefont{T.}~\bibnamefont{Lanting}},
  \bibinfo{author}{\bibfnamefont{F.}~\bibnamefont{Altomare}},
  \bibinfo{author}{\bibfnamefont{A.~J.} \bibnamefont{Berkley}},
  \bibinfo{author}{\bibfnamefont{S.}~\bibnamefont{Ejtemaee}},
  \bibinfo{author}{\bibfnamefont{E.}~\bibnamefont{Hoskinson}},
  \bibinfo{author}{\bibfnamefont{S.}~\bibnamefont{Huang}},
  \bibnamefont{et~al.}, \bibinfo{journal}{Nature Physics}
  \textbf{\bibinfo{volume}{18}}, \bibinfo{pages}{1324} (\bibinfo{year}{2022}),
  ISSN \bibinfo{issn}{1745-2481},
  \urlprefix\url{https://doi.org/10.1038/s41567-022-01741-6}.

\bibitem[{\citenamefont{Gross and Bakr}(2021)}]{Gross2021}
\bibinfo{author}{\bibfnamefont{C.}~\bibnamefont{Gross}} \bibnamefont{and}
  \bibinfo{author}{\bibfnamefont{W.~S.} \bibnamefont{Bakr}},
  \bibinfo{journal}{Nature Physics} \textbf{\bibinfo{volume}{17}},
  \bibinfo{pages}{1316} (\bibinfo{year}{2021}), ISSN \bibinfo{issn}{1745-2481},
  \urlprefix\url{https://doi.org/10.1038/s41567-021-01370-5}.

\bibitem[{\citenamefont{Pfeuty}(1970)}]{PFEUTY1970}
\bibinfo{author}{\bibfnamefont{P.}~\bibnamefont{Pfeuty}},
  \bibinfo{journal}{Annals of Physics} \textbf{\bibinfo{volume}{57}},
  \bibinfo{pages}{79} (\bibinfo{year}{1970}), ISSN \bibinfo{issn}{0003-4916},
  \urlprefix\url{https://www.sciencedirect.com/science/article/pii/0003491670902708}.

\bibitem[{\citenamefont{Schneider et~al.}(2021)\citenamefont{Schneider,
  Despres, Thomson, Tagliacozzo, and Sanchez-Palencia}}]{Schneider2021}
\bibinfo{author}{\bibfnamefont{J.~T.} \bibnamefont{Schneider}},
  \bibinfo{author}{\bibfnamefont{J.}~\bibnamefont{Despres}},
  \bibinfo{author}{\bibfnamefont{S.~J.} \bibnamefont{Thomson}},
  \bibinfo{author}{\bibfnamefont{L.}~\bibnamefont{Tagliacozzo}},
  \bibnamefont{and}
  \bibinfo{author}{\bibfnamefont{L.}~\bibnamefont{Sanchez-Palencia}},
  \bibinfo{journal}{Phys. Rev. Res.} \textbf{\bibinfo{volume}{3}},
  \bibinfo{pages}{L012022} (\bibinfo{year}{2021}),
  \urlprefix\url{https://link.aps.org/doi/10.1103/PhysRevResearch.3.L012022}.

\bibitem[{\citenamefont{Brydges et~al.}(2019)\citenamefont{Brydges, Elben,
  Jurcevic, Vermersch, Maier, Lanyon, Zoller, Blatt, and Roos}}]{Brydges2019}
\bibinfo{author}{\bibfnamefont{T.}~\bibnamefont{Brydges}},
  \bibinfo{author}{\bibfnamefont{A.}~\bibnamefont{Elben}},
  \bibinfo{author}{\bibfnamefont{P.}~\bibnamefont{Jurcevic}},
  \bibinfo{author}{\bibfnamefont{B.}~\bibnamefont{Vermersch}},
  \bibinfo{author}{\bibfnamefont{C.}~\bibnamefont{Maier}},
  \bibinfo{author}{\bibfnamefont{B.~P.} \bibnamefont{Lanyon}},
  \bibinfo{author}{\bibfnamefont{P.}~\bibnamefont{Zoller}},
  \bibinfo{author}{\bibfnamefont{R.}~\bibnamefont{Blatt}}, \bibnamefont{and}
  \bibinfo{author}{\bibfnamefont{C.~F.} \bibnamefont{Roos}},
  \bibinfo{journal}{Science} \textbf{\bibinfo{volume}{364}},
  \bibinfo{pages}{260} (\bibinfo{year}{2019}),
  \eprint{https://www.science.org/doi/pdf/10.1126/science.aau4963},
  \urlprefix\url{https://www.science.org/doi/abs/10.1126/science.aau4963}.

\bibitem[{\citenamefont{Satzinger et~al.}(2021)\citenamefont{Satzinger, Liu,
  Smith, Knapp, Newman, Jones, Chen, Quintana, Mi, Dunsworth
  et~al.}}]{Satzinger2021}
\bibinfo{author}{\bibfnamefont{K.}~\bibnamefont{Satzinger}},
  \bibinfo{author}{\bibfnamefont{Y.-J.} \bibnamefont{Liu}},
  \bibinfo{author}{\bibfnamefont{A.}~\bibnamefont{Smith}},
  \bibinfo{author}{\bibfnamefont{C.}~\bibnamefont{Knapp}},
  \bibinfo{author}{\bibfnamefont{M.}~\bibnamefont{Newman}},
  \bibinfo{author}{\bibfnamefont{N.~C.} \bibnamefont{Jones}},
  \bibinfo{author}{\bibfnamefont{Z.}~\bibnamefont{Chen}},
  \bibinfo{author}{\bibfnamefont{C.}~\bibnamefont{Quintana}},
  \bibinfo{author}{\bibfnamefont{X.}~\bibnamefont{Mi}},
  \bibinfo{author}{\bibfnamefont{A.}~\bibnamefont{Dunsworth}},
  \bibnamefont{et~al.}, \bibinfo{journal}{Science}
  \textbf{\bibinfo{volume}{374}}, \bibinfo{pages}{1237} (\bibinfo{year}{2021}),
  \urlprefix\url{https://www.science.org/doi/abs/10.1126/science.abi8378}.

\bibitem[{\citenamefont{Elben et~al.}(2023)\citenamefont{Elben, Flammia, Huang,
  Kueng, Preskill, Vermersch, and Zoller}}]{Elben2023}
\bibinfo{author}{\bibfnamefont{A.}~\bibnamefont{Elben}},
  \bibinfo{author}{\bibfnamefont{S.~T.} \bibnamefont{Flammia}},
  \bibinfo{author}{\bibfnamefont{H.-Y.} \bibnamefont{Huang}},
  \bibinfo{author}{\bibfnamefont{R.}~\bibnamefont{Kueng}},
  \bibinfo{author}{\bibfnamefont{J.}~\bibnamefont{Preskill}},
  \bibinfo{author}{\bibfnamefont{B.}~\bibnamefont{Vermersch}},
  \bibnamefont{and} \bibinfo{author}{\bibfnamefont{P.}~\bibnamefont{Zoller}},
  \bibinfo{journal}{Nature Reviews Physics} \textbf{\bibinfo{volume}{5}},
  \bibinfo{pages}{9} (\bibinfo{year}{2023}), ISSN \bibinfo{issn}{2522-5820},
  \urlprefix\url{https://doi.org/10.1038/s42254-022-00535-2}.

\bibitem[{\citenamefont{Bornet et~al.}(2024)\citenamefont{Bornet, Emperauger,
  Chen, Machado, Chern, Leclerc, G\'ely, Chew, Barredo, Lahaye
  et~al.}}]{Bornetetal2024}
\bibinfo{author}{\bibfnamefont{G.}~\bibnamefont{Bornet}},
  \bibinfo{author}{\bibfnamefont{G.}~\bibnamefont{Emperauger}},
  \bibinfo{author}{\bibfnamefont{C.}~\bibnamefont{Chen}},
  \bibinfo{author}{\bibfnamefont{F.}~\bibnamefont{Machado}},
  \bibinfo{author}{\bibfnamefont{S.}~\bibnamefont{Chern}},
  \bibinfo{author}{\bibfnamefont{L.}~\bibnamefont{Leclerc}},
  \bibinfo{author}{\bibfnamefont{B.}~\bibnamefont{G\'ely}},
  \bibinfo{author}{\bibfnamefont{Y.~T.} \bibnamefont{Chew}},
  \bibinfo{author}{\bibfnamefont{D.}~\bibnamefont{Barredo}},
  \bibinfo{author}{\bibfnamefont{T.}~\bibnamefont{Lahaye}},
  \bibnamefont{et~al.}, \bibinfo{journal}{Phys. Rev. Lett.}
  \textbf{\bibinfo{volume}{132}}, \bibinfo{pages}{263601}
  (\bibinfo{year}{2024}),
  \urlprefix\url{https://link.aps.org/doi/10.1103/PhysRevLett.132.263601}.

\bibitem[{\citenamefont{Endres et~al.}(2011)\citenamefont{Endres, Cheneau,
  Fukuhara, Weitenberg, Schauss, Gross, Mazza, Banuls, Pollet, Bloch
  et~al.}}]{Endresetal2011}
\bibinfo{author}{\bibfnamefont{M.}~\bibnamefont{Endres}},
  \bibinfo{author}{\bibfnamefont{M.}~\bibnamefont{Cheneau}},
  \bibinfo{author}{\bibfnamefont{T.}~\bibnamefont{Fukuhara}},
  \bibinfo{author}{\bibfnamefont{C.}~\bibnamefont{Weitenberg}},
  \bibinfo{author}{\bibfnamefont{P.}~\bibnamefont{Schauss}},
  \bibinfo{author}{\bibfnamefont{C.}~\bibnamefont{Gross}},
  \bibinfo{author}{\bibfnamefont{L.}~\bibnamefont{Mazza}},
  \bibinfo{author}{\bibfnamefont{M.~C.} \bibnamefont{Banuls}},
  \bibinfo{author}{\bibfnamefont{L.}~\bibnamefont{Pollet}},
  \bibinfo{author}{\bibfnamefont{I.}~\bibnamefont{Bloch}},
  \bibnamefont{et~al.}, \bibinfo{journal}{Science}
  \textbf{\bibinfo{volume}{334}}, \bibinfo{pages}{200} (\bibinfo{year}{2011}),
  \eprint{https://www.science.org/doi/pdf/10.1126/science.1209284},
  \urlprefix\url{https://www.science.org/doi/abs/10.1126/science.1209284}.

\bibitem[{\citenamefont{Bornet et~al.}(2023)\citenamefont{Bornet, Emperauger,
  Chen, Ye, Block, Bintz, Boyd, Barredo, Comparin, Mezzacapo
  et~al.}}]{Bornet_2023}
\bibinfo{author}{\bibfnamefont{G.}~\bibnamefont{Bornet}},
  \bibinfo{author}{\bibfnamefont{G.}~\bibnamefont{Emperauger}},
  \bibinfo{author}{\bibfnamefont{C.}~\bibnamefont{Chen}},
  \bibinfo{author}{\bibfnamefont{B.}~\bibnamefont{Ye}},
  \bibinfo{author}{\bibfnamefont{M.}~\bibnamefont{Block}},
  \bibinfo{author}{\bibfnamefont{M.}~\bibnamefont{Bintz}},
  \bibinfo{author}{\bibfnamefont{J.~A.} \bibnamefont{Boyd}},
  \bibinfo{author}{\bibfnamefont{D.}~\bibnamefont{Barredo}},
  \bibinfo{author}{\bibfnamefont{T.}~\bibnamefont{Comparin}},
  \bibinfo{author}{\bibfnamefont{F.}~\bibnamefont{Mezzacapo}},
  \bibnamefont{et~al.}, \bibinfo{journal}{Nature}
  \textbf{\bibinfo{volume}{621}}, \bibinfo{pages}{728} (\bibinfo{year}{2023}),
  ISSN \bibinfo{issn}{1476-4687},
  \urlprefix\url{http://dx.doi.org/10.1038/s41586-023-06414-9}.

\bibitem[{\citenamefont{Levine et~al.}(2018)\citenamefont{Levine, Keesling,
  Omran, Bernien, Schwartz, Zibrov, Endres, Greiner,
  Vuleti\ifmmode~\acute{c}\else \'{c}\fi{}, and
  Lukin}}]{PhysRevLett.121.123603}
\bibinfo{author}{\bibfnamefont{H.}~\bibnamefont{Levine}},
  \bibinfo{author}{\bibfnamefont{A.}~\bibnamefont{Keesling}},
  \bibinfo{author}{\bibfnamefont{A.}~\bibnamefont{Omran}},
  \bibinfo{author}{\bibfnamefont{H.}~\bibnamefont{Bernien}},
  \bibinfo{author}{\bibfnamefont{S.}~\bibnamefont{Schwartz}},
  \bibinfo{author}{\bibfnamefont{A.~S.} \bibnamefont{Zibrov}},
  \bibinfo{author}{\bibfnamefont{M.}~\bibnamefont{Endres}},
  \bibinfo{author}{\bibfnamefont{M.}~\bibnamefont{Greiner}},
  \bibinfo{author}{\bibfnamefont{V.}~\bibnamefont{Vuleti\ifmmode~\acute{c}\else
  \'{c}\fi{}}}, \bibnamefont{and} \bibinfo{author}{\bibfnamefont{M.~D.}
  \bibnamefont{Lukin}}, \bibinfo{journal}{Phys. Rev. Lett.}
  \textbf{\bibinfo{volume}{121}}, \bibinfo{pages}{123603}
  (\bibinfo{year}{2018}),
  \urlprefix\url{https://link.aps.org/doi/10.1103/PhysRevLett.121.123603}.

\bibitem[{\citenamefont{Kogut}(1979)}]{Kogut1979}
\bibinfo{author}{\bibfnamefont{J.~B.} \bibnamefont{Kogut}},
  \bibinfo{journal}{Rev. Mod. Phys.} \textbf{\bibinfo{volume}{51}},
  \bibinfo{pages}{659} (\bibinfo{year}{1979}),
  \urlprefix\url{https://link.aps.org/doi/10.1103/RevModPhys.51.659}.

\bibitem[{\citenamefont{Aidelsburger et~al.}(2022)\citenamefont{Aidelsburger,
  Barbiero, Bermudez, Chanda, Dauphin, Gonz\'alez-Cuadra, Grzybowski, Hands,
  Jendrzejewski, J\"unemann et~al.}}]{Aidelsburger2022}
\bibinfo{author}{\bibfnamefont{M.}~\bibnamefont{Aidelsburger}},
  \bibinfo{author}{\bibfnamefont{L.}~\bibnamefont{Barbiero}},
  \bibinfo{author}{\bibfnamefont{A.}~\bibnamefont{Bermudez}},
  \bibinfo{author}{\bibfnamefont{T.}~\bibnamefont{Chanda}},
  \bibinfo{author}{\bibfnamefont{A.}~\bibnamefont{Dauphin}},
  \bibinfo{author}{\bibfnamefont{D.}~\bibnamefont{Gonz\'alez-Cuadra}},
  \bibinfo{author}{\bibfnamefont{P.~R.} \bibnamefont{Grzybowski}},
  \bibinfo{author}{\bibfnamefont{S.}~\bibnamefont{Hands}},
  \bibinfo{author}{\bibfnamefont{F.}~\bibnamefont{Jendrzejewski}},
  \bibinfo{author}{\bibfnamefont{J.}~\bibnamefont{J\"unemann}},
  \bibnamefont{et~al.}, \bibinfo{journal}{Philosophical Transactions of the
  Royal Society A: Mathematical, Physical and Engineering Sciences}
  \textbf{\bibinfo{volume}{380}}, \bibinfo{pages}{20210064}
  (\bibinfo{year}{2022}).

\bibitem[{\citenamefont{Halimeh et~al.}(2023)\citenamefont{Halimeh,
  Aidelsburger, Grusdt, Hauke, and
  Yang}}]{halimeh2023coldatomquantumsimulatorsgauge}
\bibinfo{author}{\bibfnamefont{J.~C.} \bibnamefont{Halimeh}},
  \bibinfo{author}{\bibfnamefont{M.}~\bibnamefont{Aidelsburger}},
  \bibinfo{author}{\bibfnamefont{F.}~\bibnamefont{Grusdt}},
  \bibinfo{author}{\bibfnamefont{P.}~\bibnamefont{Hauke}}, \bibnamefont{and}
  \bibinfo{author}{\bibfnamefont{B.}~\bibnamefont{Yang}},
  \emph{\bibinfo{title}{Cold-atom quantum simulators of gauge theories}}
  (\bibinfo{year}{2023}), \eprint{2310.12201},
  \urlprefix\url{https://arxiv.org/abs/2310.12201}.

\bibitem[{\citenamefont{Gaudin}(1960)}]{Gaudin1960}
\bibinfo{author}{\bibfnamefont{M.}~\bibnamefont{Gaudin}},
  \bibinfo{journal}{Nuclear Physics} \textbf{\bibinfo{volume}{15}},
  \bibinfo{pages}{89} (\bibinfo{year}{1960}), ISSN \bibinfo{issn}{00295582}.

\bibitem[{\citenamefont{Bocini}(2023{\natexlab{a}})}]{bocini_thesis}
\bibinfo{author}{\bibfnamefont{S.}~\bibnamefont{Bocini}},
  \bibinfo{type}{Theses}, \bibinfo{school}{{Universit{\'e} Paris-Saclay}}
  (\bibinfo{year}{2023}{\natexlab{a}}),
  \urlprefix\url{https://theses.hal.science/tel-04230051}.

\bibitem[{\citenamefont{Fagotti}(2020)}]{Fagotti_2020}
\bibinfo{author}{\bibfnamefont{M.}~\bibnamefont{Fagotti}},
  \bibinfo{journal}{SciPost Phys.} \textbf{\bibinfo{volume}{8}},
  \bibinfo{pages}{048} (\bibinfo{year}{2020}),
  \urlprefix\url{https://scipost.org/10.21468/SciPostPhys.8.3.048}.

\bibitem[{\citenamefont{Bocini}(2023{\natexlab{b}})}]{Bocini_2023}
\bibinfo{author}{\bibfnamefont{S.}~\bibnamefont{Bocini}},
  \bibinfo{journal}{SciPost Physics} \textbf{\bibinfo{volume}{15}}
  (\bibinfo{year}{2023}{\natexlab{b}}), ISSN \bibinfo{issn}{2542-4653},
  \urlprefix\url{http://dx.doi.org/10.21468/SciPostPhys.15.1.027}.

\bibitem[{\citenamefont{Lieb and Robinson}(1972)}]{Lieb.Robinson1972}
\bibinfo{author}{\bibfnamefont{E.~H.} \bibnamefont{Lieb}} \bibnamefont{and}
  \bibinfo{author}{\bibfnamefont{D.~W.} \bibnamefont{Robinson}},
  \bibinfo{journal}{Commun. Math. Phys.} \textbf{\bibinfo{volume}{28}},
  \bibinfo{pages}{251} (\bibinfo{year}{1972}),
  \urlprefix\url{https://doi.org/10.1007/BF01645779}.

\end{thebibliography}

\end{document}